\documentclass[a4paper,12pt]{article}
\usepackage[a4paper, left=3cm, right=3cm, top=2cm, bottom=2cm]{geometry}
\usepackage{amsmath}
\usepackage{bm}
\usepackage{bbm}
\usepackage{amsthm}
\usepackage{dsfont}
\usepackage[utf8]{inputenc} 
\usepackage[english]{babel} 
\usepackage{xcolor} 
\usepackage{tikz}
\usetikzlibrary{positioning,arrows}
\usepackage{adjustbox} 
\usepackage[onehalfspacing]{setspace}
\usepackage{graphicx}
\usepackage{pgfplots}
\usepackage{xcolor} 
\usepackage{caption}
\usepackage{mathcomp}
\usepackage{verbatim} 
\usepackage{enumitem}
\usepackage[overload]{abraces}
\usepackage{array}
\usepackage{pict2e} 
\usepackage{amssymb}
\usepackage{epstopdf}
\usepackage{pbox}
\usepackage{mathrsfs}
\usepackage{graphicx,wrapfig,lipsum}
\usepackage{multirow}
\usepackage{subcaption}
\usepackage{colortbl}
\usepackage{mathtools}
\usepackage{multicol}
\usepackage{wasysym}
\usepackage{makecell}
\usepackage{eurosym}
\PassOptionsToPackage{hyphens}{url}
\usepackage{booktabs}

\usepackage{pdflscape}
\usepackage{afterpage}
\usepackage{booktabs,arydshln}
\usepackage{capt-of}
\usepackage{rotating}
\usepackage{longtable}
\usepackage{pdfpages}
\usepackage[hidelinks,
colorlinks=true,
urlcolor=blue,
linkcolor=blue,
citecolor=blue,
hyperfootnotes=false]{hyperref}
\usepackage{natbib}
\usepackage{mathtools}
\DeclarePairedDelimiter\ceil{\lceil}{\rceil}

\usetikzlibrary{decorations.pathreplacing}
\usetikzlibrary{shapes,arrows}

\renewcommand{\ref}[1]{\hyperref[#1]{\ref{#1}}}

\newtheorem{proposition}{Proposition}
\newtheorem{lemma}{Lemma}
\newtheorem{corollary}{Corollary}

\newtheorem{assumption}{Assumption}

\renewcommand{\thefootnote}{\fnsymbol{footnote}}
\begin{document}
\onehalfspacing
\begin{center}
{\Large \textbf{{\LARGE \textbf{Innovation Diffusion among Case-based Decision-makers}}} }\footnote{I thank three anonymous referees and an editor for their helpful comments and suggestions.}\\
\vspace{5mm}
\textit{Benson Tsz Kin Leung\footnote{Hong Kong Baptist University. Email: btkleung@hkbu.edu.hk }}\\
\textit{\today}

\end{center}

\vspace{3mm} 
\begin{abstract}
	This paper analyzes a model of innovation diffusion with case-based individuals a la \cite{gilboa1995case,gilboa1996case,gilboa1997act}, who decide whether to consume an incumbent or a new product based on their and their social neighbors' previous consumption experiences. I analyze how diffusion pattern changes with individual characteristics, innovation characteristics and social network. In particular, radical innovation leads to higher initial speed but lower acceleration compared to increment innovation. Social network with stronger overall social tie, lower degree of homophily or higher exposure of reviews from early adopters speed up diffusion of innovation.
\end{abstract}

	\vspace{5mm}

\newpage

\renewcommand{\thefootnote}{\arabic{footnote}}
\setcounter{footnote}{0}
\section{Introduction}
Understanding how innovations diffuse among consumers/producers is crucial to technological adoption in the society and the incentive of  firms to innovate. Low rate of adoption leads to forgone benefits in economic development and environmental improvement (see \cite{kemp2008diffusion}, \cite{jack2013market}). Even when the innovation is significantly better than the existing products, consumers/producers may fail to adopt it due to uncertainty, inertia and many other reasons (see \cite{stoneman1994technology}, \cite{henard2001some}). 
This paper aims to understand the diffusion patterns of new products with a simple yet rich micro-founded model. In particular, I will discuss how individual characteristics, the nature of the innovation and social network affect both short-term and long-term diffusion patterns.

An important building block of this paper is the case-based decision theory by \cite{gilboa1995case,gilboa1996case,gilboa1997act} (hereinafter GS), which I use to model adoption decisions. Roughly speaking, at each period, individuals collect cases of consumption experiences of different products from their own memory, their friends, family and their network. They normalize these experiences using their own aspiration levels, then compute a weighted sum of these experiences to aggregate the information and evaluate different products.

One key advantage of this formulation is to provide a tractable yet rich framework that, as illustrated in this paper, allows us to analyze how short-term and long-term diffusion patterns depend on consumer characteristics, social network and innovation characteristics. To the best of my knowledge, the link between diffusion and innovation characteristics is overlooked in many macro-level diffusion models (see \cite{bass1969new}, \cite{dodson1978models}, \cite{easingwood1983nonuniform}, \cite{horsky1983advertising}, \cite{van2007new} and others) and micro-founded diffusion models (see \cite{jensen1982adoption}, \cite{chatterjee1990innovation}, \cite{young2006diffusion}, \cite{campbell2013word}, \cite{frick2015innovation}, \cite{board2021learning}  and others).

The model is also especially relevant to innovation diffusion, that it assumes no knowledge requirement for individuals. Imagine an individual observes some consumption experiences of innovation from his close friends. In a Bayesian model, to infer his consumption value from the innovation, the individual has to understand what drives his close friends' consumption experiences and their decisions to adopt (or not to adopt) the innovation, and information about their close friends' close friends, etc. They also have to have prior knowledge about the quality and characteristics of the innovation. Possession of such knowledge is unrealistic especially in settings with innovations and a large population of heterogeneous individuals. 

I consider a discrete, infinite time model with an incumbent product and a new product. The game starts at period~$0$ where every individual consumes the incumbent product. It gives the incumbent product a head start and captures its informational advantage. The new product, on the other hand, enters the market in period~$1$, and is superior to the incumbent product that it yields a higher payoff for everyone. As in GS, apart from the payoff that it yields, the new product is also characterized with a similarity parameter with the incumbent product. The similarity parameter captures the extent to which individuals can use existing knowledge about the incumbent product to evaluate the new product. For example, a radical innovation induces a higher payoff than incremental innovation, but it is more difficult for individuals to use existing knowledge to evaluate, and thus has a lower similarity with the incumbent product.
In each period from period 1, each individual decides whether to consume the incumbent or the new product. After that, payoff is realized, individuals update their cases and their evaluation of the two products with everyone's consumption experiences. Lastly, they enter the next period.

The results in this paper concern the relationship between diffusion patterns and both individual and innovation characteristics, as well as social network. First, I show that if an individual consumes the new product in period $t$, he also consumes the new product for all $t'\geq t$. However, this result relies crucially on the assumption that the new product is superior for every individual. Thus, it suggests that, empirically, decisions to switch back to the incumbent product implies that the new product yields lower payoff.\footnote{This result obviously contrasts with models where the adoption decision is irreversible (e.g., \cite{campbell2013word}, \cite{frick2015innovation} and \cite{board2021learning}).} On the other hand, individuals with higher aspiration levels will adopt the new product in earlier periods, in similar spirit as in GS.\footnote{\cite{gilboa1996case} show that individuals with higher aspiration levels are more willing to experiment.} Individuals who are ``more picky" will adopt innovation earlier and the  
 diffusion pattern is characterized as a sequence of (weakly) decreasing thresholds of aspiration levels. It also shows a novel spillover effect compared to the single-agent setting in GS: individuals with low aspiration levels, who will never adopt the new product in a single-agent setting, may adopt the new product after observing early adopters' consumption experiences.

Second, some individuals might never switch to the more superior new product, which is in contrast with Bayesian models (e.g., \cite{jensen1982adoption}, \cite{frick2015innovation} and \cite{board2021learning}).\footnote{In \cite{chatterjee1990innovation} and \cite{campbell2013word}, some individuals might never adopt a product because the new product is too expensive; in \cite{chatterjee2004technology}, full adoption might not happen due to conformism.} This happens when the new innovation is only marginally superior, or when it is difficult to evaluate the new product based on knowledge of the incumbent product. Inefficient level of adoption also happens when the social network is too sparse or individuals are too heterogeneous in taste, or when individuals' aspiration levels are too low, e.g., when individuals are not ``picky" enough.

Next, I look into how diffusion patterns differ for radical versus incremental innovation, in which the former yields a higher consumption payoff but has a lower similarity with the incumbent product. 
I show that radical innovation diffuses with a higher initial speed but lower acceleration compared to incremental innovation. Intuitively, radical innovation sells better among individuals who are picky enough and willing to experiment but may fail to maintain momentum when selling to individuals who are less willing to experiment.

This result links two key concepts of diffusion curves (initial speed and acceleration) with innovation characteristics which has been, to the best of my knowledge, overlooked by the literature.\footnote{See \cite{owidtechnologyadoption} for diffusion curves of different innovation.}\textsuperscript{,}\footnote{\cite{frick2015innovation} shows how different informational environments can drive S-shape and concave diffusion curves.} It contrasts with contagion models (e.g., \cite{jackson2006diffusion}, \cite{campbell2013word} and \cite{board2021learning}) in which a larger population of early adopters also increases acceleration so there is no trade-off in initial speed and acceleration. The result also complements the existing macro-level diffusion models (see for example \cite{bass1969new}, \cite{horsky1983advertising} and \cite{bass1994bass}) which characterize the diffusion pattern using reduced-form differential equations. It is in general difficult, if not impossible, to pin down the different parameters of differential equations with limited (or without any) data.\footnote{See \cite{van1997bias} and \cite{bemmaor2002impact}.} Given the heterogeneity of estimated parameters in the existing literature (\cite{sultan1990meta}), it is difficult to apply existing estimation of macro-level diffusion models during early periods of or before new product launch. The result of this paper informs decision makers on how their innovation would diffuse in the market and the welfare implications using only information of the product characteristics, e.g., how ``radical" their innovation is. 

Lastly, I analyze how social network affects product diffusion. Increased connectivity in the network speeds up diffusion when the increase in network strength is not corrected with aspiration level. Diffusion is slower if network ties with individuals with low aspiration levels, i.e., later adopters, are stronger, and is faster if individuals are more connected with individuals with high aspiration levels. It implies that platform strategies that highlight reviews from consumers who experiment more can speed up adoption and encourage innovation.

The paper is outlined as follows. I introduce the model setup in the next section. In Section 3, I show that diffusion patterns can be characterized by a decreasing sequence of aspiration thresholds. Section 4 compares the diffusion patterns with different innovation characteristics, while Section 5 presents results on how social network affects diffusion. Lastly, I conclude. 

\section{Model Setup}
Consider a market where $N$ individuals decide whether to consume one of two products. One of the two products is an incumbent product, denoted as $p_c$, and the other is a new product, denoted as $p_n$. Time is discrete and runs from $t=0,1,2,\cdots$. In period~$0$, only $p_c$ is available. Individuals, indexed with $i=1,\cdots, N$, thus all consume the incumbent product in period $t=0$. In each period $t=1,2,\cdots$, they decide whether to consume $p_c$ or $p_n$, thus switching to the new product is reversible. We denote that set of individuals who consume $p_c$ and $p_n$ in period $t$ as $D_c^t$ and $D_n^t$, which with some abuse of notations also denote the number of individuals consuming the two products in period $t$. Obviously, $D_c^t\cap D_n^t=\emptyset$ and $D_c^t \cup D_n^t=\lbrace 1,2,\cdots, N\rbrace$ for all $t$, and by assumption $D_n^0=\emptyset$. In the following, we analyze how $D_c^t$ and $D_n^t$ evolves with time.

\paragraph{Individuals} 
Each individual $i$ is endowed with cases $C_{t}$ in period $t$, in which its elements are tuple $(j,p,v)\in \lbrace 1,\cdots,N\rbrace\times \lbrace p_c,p_n\rbrace\times \mathbb{R}$. Each case $(j,p,v)$ is a review/case of consumption, which says an individual $j$ consumed a product $p$ and got payoff $v$. Following GS, a product $p$ is evaluated with the following $U$ function:
\begin{equation}\label{eq:ufunction}
U_{i}^t(p)=\sum_{(j,p',v)\in C_t}s_{i,j}s_{p,p'}(v-H_{i}).
\end{equation}
I call $U_{i}^t(p)$ as the individual $i$'s evaluation of product $p$ given cases $C_t$. Using the terminologies in GS, $H_{i}$ is individual $i$'s aspiration level, and $s_{i,j}$ and $s_{p,p'}$ are called the similarity between individuals and products, respectively. Equation~\eqref{eq:ufunction} could be interpreted as follows: in period~$t$ individuals collect a database of consumption experience $C_t$ from their previous selves and others, and then aggregate different individuals' consumption experiences with different products using the weights $s_{i,j}s_{p,p'}$, while normalizing the experiences using their own aspiration level $H_i$.\footnote{Another functional form, inspired by the average similarity function of \cite{gilboa1996case}, is:
\begin{equation*}
	U_{i}^t(p)=\frac{\sum_{(j,p',v)\in C_t}s_{i,j}s_{p,p'}(v-H_{i})}{\sum_{(j,p',v)\in C_t}s_{i,j}}.
\end{equation*} That is, the individual averages his neighbors' experiences to evaluate different products. Under this formulation, all results except Proposition~\ref{prop:speedvsacceleration} qualitatively hold. I will discuss in more detail the departure after I present Proposition~\ref{prop:speedvsacceleration}. I thank an anonymous referee for this suggestion.}

In the baseline model, for simplicity, I further assume that the similarity between individuals follows (this assumption will be relaxed in Section~\ref{sec:network}):
\begin{equation}\label{eq:assumptionij}
s_{i,j}=\begin{cases}
	s \in [0,1]\text{ if $i\neq j$;}\\
	1 \text{ if $i=j$.}
\end{cases}
\end{equation}
The assumption regarding to the similarity between products $s_{p,p'}$ will be discussed later when I introduce the products.  Equation~\eqref{eq:assumptionij} implies that individuals discount consumption experiences from others. Note that $s_{i,j}$ could also be interpreted as a weighted and directed network among the $N$ individuals. In this case, Equation~\eqref{eq:assumptionij} implies that the the social network among the $N$ individuals is a symmetric and complete network where the strength of cross-individuals ties are $s$ and the strength of self connection is $1$. Depending on the interpretation, the parameter $s$ measures the diversity of individuals or the sparsity of the social network. When individuals are diverse, e.g., they have very different taste towards different attributes of the product, or when their network ties are weak, experiences of others have small influence on one's evaluation of the product.

On the other hand, individuals are heterogeneous in their aspiration levels $H_{i}\in\mathbb{R}$. Without loss of generality and for the ease of exposition, I assume that there are $G\leq N$ mutually-exclusive groups of individuals with different aspiration levels, with $N_{1},N_{2},\cdots, N_{G}$ denoting the set itself and also its cardinality. $H_{i}$ differs only across group such that $H_{i}=H_{j}$ if and only if $i,j\in N_{k}$ for some $k=1,\cdots,G$. With some abuse of notations, I also denote the aspiration levels of individuals in a group $N_{k}$ as $H_{N_{k}}$. Without loss of generality, I order the aspiration levels such that $H_{N_{k}}$ is strictly decreasing in $k$.

\paragraph{Products} 
In period~$0$, only the incumbent product is available in the market. The incumbent product $p_c$ generates payoff $v=v_{L}$ for some constant $v_{L}\in (H_{N_{2}},H_{N_{1}}]$ for each consumption.\footnote{The assumption where $H_{N_{1}}\geq v_{L}$ ensures that at least some individuals will adopt the new product at some period~$t$. On the other hand, the assumption that $H_{N_{2}}< v_{L}$ induces no loss of generality as $H\geq v_{L}$ does not affect the diffusion pattern: all individuals with $H\geq v_{L}$ adopt the new product in period 1 onward and the evaluations of all other individuals do not depend on the aspiration levels of individuals who have $H\geq v_{L}$.}\textsuperscript{,}\footnote{For ease of exposition, I assume that the payoff generated by the incumbent product, $v_{Li}$, does not vary across $i$. The results in this paper holds when this assumption is relaxed, as long as $\min_{i}{v_{Hi}}>\max_i v_{Li}$. } 

A new product, denoted as $p_n$, is introduced to the market at period $1$. The payoff generated by the new product is ex-ante uncertain, could be different for different individuals but is stable across periods. In the following analysis, I fix a particular realization of payoff $v_{Hi}$ for $i=1,\cdots, N$, and analyze the resulting adoption dynamics.  I assume that the new product is superior to the incumbent product for all individuals such that $v_{Hi}\geq v_L$ and $v_{Hi}\geq H_{N_1}$ for all individual $i$.\footnote{Note that $v_{Hi}\geq H_{N_1}$ implies $v_{Hi}\geq v_L$ for all $i$. The assumption that $v_{Hi}\geq H_{N_1}$ for all $i$ ensures that individuals in group $N_1$ is satisfied with the new product and thus does not switch back to the incumbent product. If the size of group $N_1$ is small enough, our results continue to hold qualitatively for all other groups' adoption pattern, e.g., individuals in $N_1$ might switch between the incumbent and the new product, but individuals in other groups will never switch back to the incumbent product.} Therefore, everyone switching to the new product is the socially desirable outcome. I make the following assumption for the similarity between the two products.

\begin{assumption}
The similarity function is asymmetric, such that $s_{p_{n},p_{c}}\neq s_{p_{c},p_{n}}$. Moreover, $1>s_{p_{n},p_{c}}=s_{p}> s_{p_{c},p_{n}}=0$ for some constant $s_p$ and $s_{p_{n},p_{n}}=s_{p_{c},p_{c}}=1$.
\end{assumption}
This assumption 
captures the differences in learning about an established versus a new product. Briefly speaking, the asymmetry in the similarity function $s_{p_{n},p_{c}}> s_{p_{c},p_{n}}$ implies there are more learning for the new product than the incumbent product.\footnote{The same insight holds for example in Bayesian updating with Gaussian distributed state and signal. As an example, with state $\omega\sim N(0,\sigma_{\omega}^2)$ and signal $x\sim N(x,\sigma_{x}^2)$, the posterior expectation $E(\omega\vert x)$ is equal to $\frac{\sigma_{\omega}^2}{\sigma_{x}^2+\sigma_{\omega}^2}x$. The more uncertain the state is, i.e., the larger $\sigma_{\omega}^2$, the more sensitive the expected state is to the signal. If in contrast $s_{p_n,p_c\leq s_{p_c,p_n}}$, no individuals will ever adopt the new product.} As will be clear in the following sections, what matters to product diffusion is the ratio $\frac{1-s_{p_{c},p_{n}}}{1-s_{p_{n},p_{c}}}$. Thus, without loss of generality I normalize $s_{p_{c},p_{n}}=0$. I call $((v_{Hj})_{j=1}^{N},s_p)$ the (realized) product specification of the new product.

To focus on the impact of individual and product characteristics on diffusion, I assume both products are sold at the same price in each period which is normalized as $0$.\footnote{Thus, firms make no decisions and I analyze how the product diffuses among the $N$ individuals. However, the results on diffusion patterns, e.g., how product characteristics affect diffusion, inform about firms' decisions which we will discuss in section~\ref{sec:product}. }

\paragraph{Cases and Timeline}
I now outline the evolution of cases endowed by individuals and the timeline of the game. At $t=0$, $C_0=\emptyset$ and all individuals consume the incumbent product $p_c$. 
In each period $t=1,\cdots,\infty$, each individual is endowed with $C_t$ and consume the product $p$ that has the higher evaluation $U_{i}^t(p)$ and the incumbent product if $U_{i}^t(p_c)=U_{i}^t(p_n)$. 
The cases at any period $t$, $C_t$, is the union of the cases in the previous period $C_{t-1}$ and the cases generated by the consumption in period $t-1$. That is, $C_t=C_{t-1}\cup (j,p_{c},v_{L})_{j\in D_{c}^{t-1}}\cup (j,p_{n},v_{Hj})_{j\in D_{n}^{t-1}}$.

\paragraph{Discussions} In the following, I discuss the assumptions of the model and in particular the use of case-based decision theory.

\begin{enumerate}[leftmargin=*]
	\item \textbf{Interpretation.} First, case-based decision theory offers a nice and intuitive interpretation on individuals' thinking process of product adoption. More specifically, cases could be interpreted as reviews that are posted online, or just experiences transmitted among a social network. $s_{i,j}$ thus captures how reviews or in general information transmits in the social network among the $N$ individuals, while the $U$ function represents a simple way of aggregating those information.\footnote{Interested reader could refer to \cite{gilboa1995case,gilboa1996case,gilboa1997act} for the axiomatic foundations of the case-based decision theory.}
	
	
	\item \textbf{No prior knowledge about the new product.} Second, and importantly, the case-based decision theory does not assume prior knowledge of individuals about the new product. In contrast, Bayesian learning assumes that individuals are endowed with a prior belief over the payoffs generated by the new product, and information structure that generates other's consumption experiences. Processing such prior knowledge is especially unrealistic in settings of new innovations and when the heterogeneity of individual tastes is non-trivial. In contrast, cases-based decision theory assumes that individuals use an evaluation algorithm that requires no prior knowledge.
\end{enumerate}

\section{Adoption Dynamics}
In this section, I present the diffusion dynamics, i.e., the individuals' consumption in each period, given a (realized) product specification $((v_{Hj})_{j=1}^{N},s_p)$ and individual characteristics $s$ and $H_{N_{1}},\cdots,H_{N_{G}}$. The following Lemma shows that although adoption of the new product is reversible, individuals do not switch back to the incumbent product. 



\begin{lemma}\label{lemma_switch}
Once an individual chooses the new product in period~$t$, he chooses the new product in all future periods. Thus, $D_{n}^{t-1}\subseteq D_{n}^{t}$ for all $t$.
\end{lemma}

The proof of Lemma~\ref{lemma_switch} and other results are shown in the Appendix. The Lemma is proved by induction. If an individual $i$ switches from the incumbent product to the new product in period $t$, it must be that the consumption cases in period $t-1$ increases the evaluation of the new product more than that of the incumbent product. Next, in period $t$, as there are more individuals, including $i$, consuming the new product, the evaluation of the new product must increase even more than the increase of the evaluation of the incumbent product. Therefore, individual $i$ also consumes the new product in period $t+1$.

Note that Lemma~\ref{lemma_switch} is driven by the assumption that the new product is superior to the incumbent product, i.e., $v_{Hi}>v_L$ for all $i$. If in contrast $v_{Hi}<v_L$ for some $i$, then individuals (not restricted to $i$) may switch back to the incumbent product after adopting the new product. The larger $v_L-v_{Hi}$ is, the more likely that some individuals will switch back to the incumbent product. Such switchback rate is thus an indicator of the quality of the innovation. As an example, after its launch in 2011, the social media platform Google+ has seen a  ``98\% year-over-year" decrease in engagement rate and is eventually shut down in 2019.\footnote{See \url{https://www.forbes.com/sites/stevedenning/2015/04/23/has-google-really-died/?sh=2d2c9f65466c} and  \url{https://www.theverge.com/2019/4/2/18290637/google-plus-shutdown-consumer-personal-account-delete}.}


The following Proposition further shows that the population of individuals that buy the new product in period $t$ must be a weakly increasing segment of aspiration levels.

\begin{proposition}\label{prop_segments}
In each period, there exists a unique threshold $\underline{H}_{t}\in\mathbb{R}\cup\lbrace-\infty\rbrace$ such that all and only individuals whose aspiration level $H_{i}> \underline{H}_{t}$ consume the new product in period $t$. Moreover, $\underline{H}_{t}$ is weakly decreasing in $t$.
\end{proposition}

The adoption process is thus characterized by a unique sequence of (weakly) decreasing thresholds $\underline{H}_{1},\underline{H}_{2},\cdots$.\footnote{When the payoff from the incumbent product $v_{Li}$ varies across $i$, individuals with lower $v_{Li}-H_i$ adopt the new product earlier. } As in the spirit of GS, individuals with higher aspiration levels are more inclined to experiment and adopt the new product: the higher $H$ is, the higher the ratio $\frac{v_{Hi}-H}{v_{L}-H}$ is, and the higher the increase of the evaluation of new product relative to the incumbent product is. Different from GS, the current multi-agents model additionally gives rise to a novel spillover effect. In a single-agent setting, only individuals in $N_1$ will ever adopt the new product as $H_{N_1}>v_L$ and $s_p<1$.\footnote{If the two inequalities do not hold, no individuals will ever adopt the new product.} In contrast, here even individuals whose aspiration levels are so low that they would stick with the incumbent product in a single-agent setting may eventually switch to the new product, because they observe experiences of individuals with higher aspiration levels. As an example to illustrate the importance of this spillover effect, in the early days of Reddit, their founders created fake accounts,  content and comments to show how lively their website is, and thus acted themselves as ``early adopters" to encourage ``late adopters".\footnote{See \url{https://arstechnica.com/information-technology/2012/06/reddit-founders-made-hundreds-of-fake-profiles-so-site-looked-popular/}.}

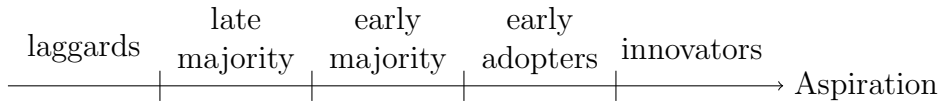
\begin{figure}
	\begin{center}
		\begin{tikzpicture}
			\draw[->] (0,0)--(10.2,0) node[right] {Aspiration};
			\draw (2,0.2)--(2,-0.2);
			\draw (4,0.2)--(4,-0.2);
			\draw (6,0.2)--(6,-0.2);
			\draw (8,0.2)--(8,-0.2);
			\node at (1,0.5) {laggards};
			\node[align=center] at (3,0.6) {late \\majority};
			\node[align=center] at (5,0.6) {early \\majority};
			\node[align=center] at (7,0.6) {early \\adopters};
			\node at (9,0.5) {innovators};
		\end{tikzpicture}
	\end{center}
	\caption{Classification of individuals a la \cite{rogers2010diffusion} according to their aspiration levels.}
	\label{fig:rogers}
\end{figure}

 Not only Lemma~\ref{lemma_switch} and Proposition~\ref{prop_segments} provide a simple characterization of the diffusion pattern, they also show a link between individuals' willingness to adopt new products and how much payoff they aspire for. It establishes a testable implication and a correspondence between this micro-based diffusion model and the macro-based diffusion model like \cite{rogers2010diffusion}, e.g., individuals are classified into innovators, early adopters, early majority, late majority and laggards in a descending order of aspiration levels, as illustrated in Figure~\ref{fig:rogers}.

Now, for ease of notations, denote $F(H)$ as the complementary cumulative mass function of aspiration levels, i.e., the mass of individuals with aspiration level strictly higher than $H$:
\begin{equation*}
F(H)=\sum_{n=1}^{k}N_{n}\text{ where $H_{N_{k+1}}\leq H<H_{N_{k}}$}.
\end{equation*}
Also denote $v_{>H}$ as the average aspiration level given that it is strictly higher than $H$, i.e., $v_{>H}=\frac{\sum_{i: v_{Hi}>H}v_{Hi}}{F(H)}$. The following Proposition presents the results on the asymptotic market share of the new product.

\begin{proposition}\label{prop:coverage}
All individuals eventually consume the new product if and only if there does not exist a $H\in[H_{N_{G}},H_{N_{1}})$ such that
\begin{equation}\label{eq:coverage1}
\dfrac{v_{>H}-v_L}{v_{L}-H}\leq \dfrac{(1-s_{p})\left[s[N-F(H)-1]+1\right]}{s[F(H)]}.
\end{equation}
Otherwise, only individuals in $N_{1},\cdots,N_{\bar{G}-1}$ eventually consume the new product, where $\bar{G}$ follows:
\begin{equation*}
	\bar{G}=\min k \text{ where } \dfrac{v_{>H_{N_{k}}}-H_{N_{k}}}{v_{L}-H_{N_{k}}}\leq \dfrac{(1-s_{p})\left[s[F(H_{N_{k}})-1]+1\right]}{s[N-F(H_{N_{k}})]},
\end{equation*}  
while the individuals in $N_{\bar{G}},\cdots,N_{G}$ consume the incumbent product asymptotically. Moreover, fixing the aspiration levels $H_{N_{1}},\cdots,H_{N_{G}}$, when $F(H)$ increases for all $H$, $\bar{G}$ increases.
\end{proposition}

The asymptotic market share of the new product sheds light on the upper bound of social efficiency, and also contrasts with existing Bayesian models where all individuals adopt the more superior new product asymptotically.\footnote{For example, see \cite{jensen1982adoption}, \cite{frick2015innovation} and \cite{board2021learning}. In Bayesian models, individuals eventually learn that the new product is more superior and adopt the new product.} It corresponds to situation where experiences propagate in the society quickly such that the duration between periods is small and future payoffs are not discounted. The next two sections will shed light on situations where future payoffs are discounted. Note that Inequality~\eqref{eq:coverage1} is more likely to fail when $H_{N_G}$, $s$, $(v_{Hi})_{i=1}^{N}$ or $s_{p}$ is large, thus Proposition~\ref{prop:coverage} presents intuitive examples where there is no asymptotic inefficiency: when individuals have high aspiration levels and thus are more willing to experiment; when experiences of individuals has big impact on other individuals' evaluation, for example when they have similar taste or strong social ties; when the new product is much more superior; or when it is easy to use knowledge of the incumbent product to evaluate the new product. Also, unsurprisingly, the last part of  Proposition~\ref{prop:coverage} shows that when more individuals have higher aspiration levels, the new product covers a larger market asymptotically, i.e., there is a smaller asymptotic inefficiency.

\section{Comparison of Product Specifications}\label{sec:product}
Given the characterization of the adoption process, I now analyze the role of the characteristics of the new product. More specifically, I compare the diffusion process of two product specifications, $((v_{Hi})_{i=1}^N,s_{p})$ and $((v'_{Hi})_{i=1}^N,s'_{p})$, fixing other model parameters. According to Proposition~\ref{prop_segments}, the two product specifications induce two adoption processes that are characterized by two sequences of thresholds. Denote the two sequences of thresholds by $\underline{H}_{1},\underline{H}_{2},\cdots$ and $\underline{H}'_{1},\underline{H}'_{2},\cdots$ respectively.

The following Lemma shows that increasing $(v_{Hi})_{i=1}^N$ (resp. $s_{p}$) while keeping fixed $s_{p}$ (resp. $(v_{Hi})_{i=1}^N,$) decreases $\underline{H}_{t}$ for all $t\geq 2$, i.e., the product with higher quality $v_{H}$ or similarity $s_{p}$ diffuses faster in the market.
\begin{lemma}\label{lemma:dominating}
Consider two product specifications $((v_{Hi})_{i=1}^N,s_{p})$ and $((v'_{Hi})_{i=1}^N,s'_{p})$ where $((v'_{Hi})_{i=1}^N,s'_{p})\geq ((v_{Hi})_{i=1}^N,s_{p})$. For all $t\geq 2$, we have $\underline{H}_{t}\geq\underline{H}'_{t}$ for all $t\geq 2$, i.e., the product $((v'_{Hi})_{i=1}^N,s'_{p})$ covers a larger market than $((v_{Hi})_{i=1}^N,s_{p})$ in all period $t\geq 2$. 
\end{lemma}

Lemma~\ref{lemma:dominating} shows that a product specification $((v'_{Hi})_{i=1}^N,s'_{p})$ dominates another product specification $((v_{Hi})_{i=1}^N,s_{p})$ if $((v'_{H})_{i=1}^N,s'_{p})\geq ((v_{H})_{i=1}^N,s_{p})$, i.e., the former covers a larger market than the latter in each period $t\geq 2$. The following Proposition looks into the more interesting case where no product specification dominates the other, i.e., one product covers a larger market in some periods but covers a smaller market in other periods.

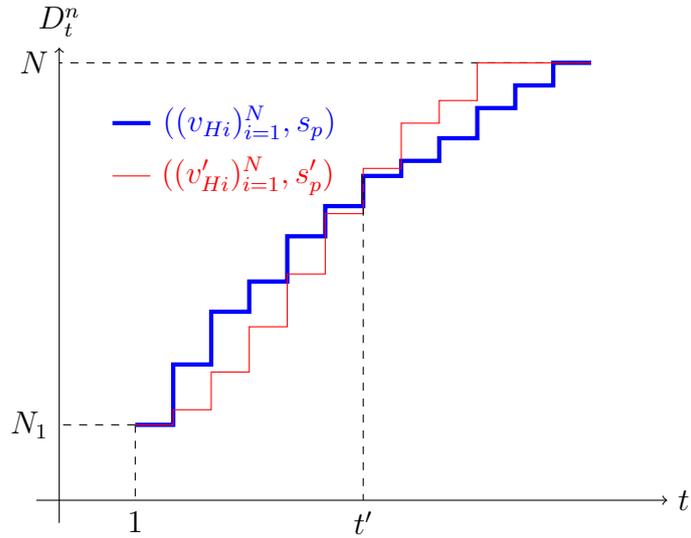
\begin{figure}
	\begin{center}
		\begin{tikzpicture}
			\draw[->] (0,-0.3)--(0,6) node[above] {$D^{n}_{t}$};
			\draw[->] (-0.3,0)--(8,0) node[right] {$t$};
			\draw[dashed] (1,1) -- (1,-0) node[below] {$1$};
			\draw[dashed] (1,1) -- (0,1) node[left] {$N_{1}$};
			\draw[ultra thick, blue] (1,1) -- (1.5,1)--(1.5,1.8)--(2,1.8)--(2,2.5)--(2.5,2.5)--(2.5,2.9)--(3,2.9)--(3,3.5)--(3.5,3.5)--(3.5,3.9)--(4,3.9)--(4,4.3)--(4.5,4.3)--(4.5,4.5)--(5,4.5)--(5,4.8)--(5.5,4.8)--(5.5,5.2)--(6,5.2)--(6,5.5)--(6.5,5.5)--(6.5,5.8)--(7,5.8);
			\draw[dashed] (7,5.8)--(0,5.8) node[left] {$N$};
			\draw[red] (1,1) -- (1.5,1)--(1.5,1.2)--(2,1.2)--(2,1.7)--(2.5,1.7)--(2.5,2.3)--(3,2.3)--(3,3)--(3.5,3)--(3.5,3.8)--(4,3.8)--(4,4.4)--(4.5,4.4)--(4.5,5)--(5,5)--(5,5.3)--(5.5,5.3)--(5.5,5.8)--(6,5.8)--(6,5.8)--(6.5,5.8)--(6.5,5.8)--(7,5.8);
			\draw[dashed] (4,4.3)--(4,0) node[below] {$t'$};
			\draw[ultra thick, blue] (0.7,5)--(1.2,5) node[right] {$((v_{Hi})_{i=1}^{N},s_{p})$};
			\draw[red] (0.7,4.3)--(1.2,4.3) node[right] {$((v'_{Hi})_{i=1}^N,s'_{p})$};
		\end{tikzpicture}
	\end{center}
	\caption{An example of diffusion patterns of two non-dominated product specifications $((v_{Hi})_{i=1}^{N},s_{p})$ and $((v'_{Hi})_{i=1}^N,s'_{p})$ where $v_{Hi}=v_{H}>v'_{H}=v'_{Hi}$ for all $i$ but $s_{p}<s'_{p}$. Product $((v_{Hi})_{i=1}^{N},s_{p})$ diffuses faster at the beginning but slower as $t$ grows larger.}
	\label{fig:diffusioncomparison}
\end{figure}

\begin{proposition}\label{prop:speedvsacceleration}
Consider two product specifications $((v'_{Hi})_{i=1}^N,s'_{p})$ and $((v_{Hi})_{i=1}^N,s_{p})$ where $v_{Hi}=v_H$, $v'_{Hi}=v'_H$ for all $i$, and $(v'_{H}-v_{H})(s'_{p}-s_{p})<0$. Suppose $\underline{H}_{2}<\underline{H}'_{2}$.\footnote{The condition holds when $\frac{1}{1-s_p}\frac{v_H-\underline{H}'_2}{v_L-\underline{H}'_2}>\frac{1}{1-s'_p}\frac{v'_H-\underline{H}'_2}{v_L-\underline{H}'_2}=\frac{s(2N-N_1-2)+2}{sN_1}$, for example when $v_H$ is big enough compared to $v'_H$ or when $s_p$ is big enough compared to $s'_p$.}\textsuperscript{,}\footnote{Note that when $\underline{H}_{2}=\underline{H}'_{2}$, the second bullet point of the Proposition continues to hold. On the other hand, for the first bullet point, we have if $s_{p}> s'_{p}$, $\underline{H}_{t}\leq \underline{H}'_{t}$ for all $t\geq 3$. Moreover, if $s_{p}> s'_{p}$ and $N_1\subset D_n^{t'}$ for some $t'$, there exists $\underline{t}>t'$ such that $\underline{H}_{t}< \underline{H}'_{t}$ for all $t\geq \underline{t}$ when at least one of $\underline{H}_{t}$,$\underline{H}'_{t}$ is not $-\infty$.} We have
\begin{enumerate}
	\item if $s_{p}> s'_{p}$, as long as at least one of $\underline{H}_{t}$,$\underline{H}'_{t}$ is not $-\infty$, $\underline{H}_{t}<\underline{H}'_{t}$ for all $t\geq 3$;
	\item if $s_{p}< s'_{p}$, as long as at least one of $\underline{H}_{t}$,$\underline{H}'_{t}$ is not $-\infty$, either,
	\begin{enumerate}
		\item $\underline{H}_{t}\leq\underline{H}'_{t}$ for all $t\geq 3$,  or;
		\item there exists a $\tilde{t}\geq 2$ such that $\underline{H}_{t}\leq\underline{H}'_{t}$ if and only if $t\leq \tilde{t}$.
	\end{enumerate}
\end{enumerate}
	
\end{proposition}
The Proposition focuses on the case where all individuals get the same payoff from the new product, but because of continuity, the result also holds when $v_{Hi}$ does not vary a lot across $i$.\footnote{As an extreme counter-example, consider case 2b of the Proposition where $((v_{Hi})_{i=1}^N,s_{p})$ sells to more individuals for all period $t\leq\bar{t}$ and sell to fewer individuals at period $\bar{t}+1$. Now suppose in period $\bar{t}+1$ some individual $j$ adopts the new product $((v_{Hi})_{i=1}^N,s_{p})$ with $v_{Hj}>>v'_{Hj}$. Such consumption case will boost a lot the evaluation of $((v_{Hi})_{i=1}^N,s_{p})$  such that it sells more than $((v'_{Hi})_{i=1}^N,s'_{p})$  in period $\bar{t}+2$.} Proposition~\ref{prop:speedvsacceleration} shows, for non-dominated pair of product specifications,  an interesting difference in their diffusion patterns. The result is illustrated in Figure~\ref{fig:diffusioncomparison}. More specifically, borrowing terminologies in  \cite{bass1969new}, the diffusion patterns of the two product specifications differ in their initial speed and acceleration: the product with higher quality $v_{H}>v'_{H}$ enjoys a higher initial speed of diffusion but a lower acceleration compared to the product with higher similarity $s'_{p}>s_{p}$. 

The intuition is as follows. As shown in Proposition~\ref{prop_segments}, at the beginning of the diffusion process, individuals with high aspiration levels, i.e., innovators or early majorities a la \cite{rogers2010diffusion}, adopt the new product. Those individuals value high quality more than individuals with low aspiration levels: the marginal increase in $\frac{v_{H}-H}{v_{L}-H}$ w.r.t.~$v_H$ increases in $H$. As a result, products with higher quality attract early adopters better than products with higher similarity and thus covers a larger market at the beginning of the diffusion process. In contrast, at the later stage of the diffusion process, the adoption decisions of individuals with lower aspiration levels matter. Those individuals value high quality less and will only adopt the new product when there is sufficient knowledge of the new product in the market: their decision depends more on the reduction in ``ambiguity" than the quality of the product. Thus, as the product with higher similarity enjoys a larger knowledge spillover from the incumbent product, it sells better to individuals with lower aspiration levels and thus covers a larger market in the later stage of diffusion.\footnote{Note that the heterogeneity in aspiration levels is crucial to this trade-off between initial speed and acceleration. In particular, when $v_H>v'_H$, $\frac{v_H-H}{v_L-H}/\frac{v'_H-H}{v_L-H}$ increases in $H$: individuals with higher $H$ appreciate the product with higher payoff better. In contrast, although the heterogeneity in payoff $v_{Li}$ drives a similar diffusion process in which individuals with lower $v_{Li}-H_i$ adopt the new product earlier, individual with lower $v_{Li}$ does not appreciate the product with higher payoff better:  $\frac{v_H-H}{v_{Li}-H}/\frac{v'_H-H}{v_{Li}-H}$ is invariant in $v_{Li}$.}

This novel trade-off between initial speed and acceleration contrasts with the existing literature and offers important economic implications. First, in the existing literature of network contagion model, for example word-of-mouth model (\cite{campbell2013word}), or model where individuals adopt if a sufficient proportion of their neighbors adopts (\cite{jackson2006diffusion}), there is no trade-off between initial speed and acceleration of diffusion. If product A sells to more individuals at the beginning than product B, those initial sales help the diffusion of the product and thus product A will also enjoy a higher acceleration of diffusion than product B.\footnote{Intuitively, the argument also applies in Bayesian models such that the trade-off between initial speed and acceleration of diffusion is absent. Assuming no price different between two superior new products, if a product diffuses faster at the beginning, it generates more favorable information and thus also enjoys a higher acceleration of diffusion. The trade-off is also absent under the re-scaled formulation mentioned in page 5 footnote 7: if a product covers a larger market at the beginning, it covers a larger market for all $t$.} Moreover, these contagion model predicts that diffusion is slow at the beginning and pick up the pace once a substantial amount of people has adopted new product. In contrast, Proposition~\ref{prop:speedvsacceleration} shows that products with higher $v_H$ and low $s_p$ diffuse faster at the beginning and slow down eventually, which is documented by recent empirical study (\cite{jin2019emergence}).

Second, different from the macro-diffusion model, for example the Bass model (\cite{bass1969new}) or different epidemic models (\cite{pastor2015epidemic}), Lemma~\ref{lemma:dominating} and Proposition~\ref{prop:speedvsacceleration} characterize the relationship between innovation characteristics and diffusion patterns, thus provide a richer environment and predictions. The result suggests that radical innovation, i.e., product with high quality $v_{H}$ but low similarity $s_{p}$, diffuses quicker among consumers who are more willing to experiment but could fail to maintain momentum in the later stage of diffusion. On the other hand, incremental innovation, i.e., product with low quality $v_{H}$ but high similarity $s_{p}$, diffuse slower at the beginning but are better at maintaining momentum.\footnote{The following example illustrates why radical (resp.~incremental) innovation corresponds to a high (resp.~low) $v_H$ but low (resp.~high) $s_p$. Consider a product that is composed of multiple features. Replacing each feature with a new technology increases consumption payoff but decreases similarity. Radical innovation corresponds to a product design where many features are replaced, thus gives high payoff but has a low similarity with the original product. The opposite holds true for incremental innovation.} The result helps firms or social planners to conduct welfare analysis when there is a lack of data to estimate the existing macro-diffusion model. 
In particular, firms or social planners with less patience would prefer radical innovations than incremental innovations.


\section{Social Network and Diffusion}\label{sec:network}
In the section, I analyze and discuss the effect of social network, i.e., $(s_{i,j})$, on product diffusion. 

First, moving away from the assumption that $s_{i,i}=1$ and $s_{i,j}=s$ for all $i\neq j$, Lemma~\ref{lemma_switch} continues to hold. That is, individuals never consume the incumbent product after adopting the new product. However, Proposition~\ref{prop_segments} may not hold. In particular, an individual with lower aspiration level will adopt the new product earlier if he is better connected to individuals that adopt the new product in early periods (for example individuals in $N_1$). In the following, I show in two simple settings how homophily and stronger social ties affect diffusion.

\subsection{Homophily}
I first analyze how homophily, i.e., the tendency of individuals to form closer links with others who are similar to themselves, affects product diffusion. For simplicity, I model homophily with the following assumption on $s_{i,j}$:
\begin{equation*}
	s_{i,j}=\begin{cases}
		1&\text{ when $i=j$};\\
		\gamma s&\text{ when $i\neq j$ and $i,j\in N_{k}$ for some $k$};\\
		 s &\text{ otherwise}.
	\end{cases}
\end{equation*}
where $\gamma\in(0,\frac{1}{s})$. An increase in $\gamma$ captures an increase in homophily. Furthermore, for simplicity, I assume that all groups of individuals are of the same size, i.e., $N_{i}=N_{j}=\frac{N}{G}$ for all $i,j=1,\cdots, G$.\footnote{When different groups of individuals have different size, under this setting of homophily, there is also a group size effect as homophily has a bigger effect on consumers from a larger group: individuals from a larger group will have less incentive to adopt the new product. As a result, individuals with higher aspiration level and also a larger group size may not adopt the new product in earlier period.}

\begin{proposition}\label{prop:homophily}
The diffusion process is characterized by a sequence of decreasing threshold $\underline{H}_t\in\mathbb{R}\cup\lbrace-\infty\rbrace$ such that all and only individuals whose aspiration level $H_i>\underline{H}_t$ consumes the new product in period $t$. An increase in $\gamma$ increases $\underline{H}_{t}$ for all $t\geq 2$, i.e., an increase in homophily slows down the diffusion process.
\end{proposition}

The intuition is as follows. First note that individuals in the same group always adopt the new product at the same time. When an individual in group $N_{k}$ has not yet adopted the new product, it implies that other individuals in the same group have only consumed the incumbent product in all previous periods. As a result, an increase in homophily implies that individuals take in account more experiences of the incumbent product before they adopt the new product, which slows down the diffusion process of the new product.

\subsection{Network ties based on Aspiration Level}
Next I look into the case where $s_{i,j}$ are invariant across all $i$ and $j\in N_k$ for some $k$, and $s_{i,i}=1$ for all $i$. That is, the strength of the connection between $i$ and $j$ only depends on which group $j$ is in. For the ease of exposition, $s_{N_k}$ denotes $s_{i,j}$ where $j\in N_k$, and I call $(s_{N_k})_{k=1}^G$ a social network. Note that different from previous sections, the diffusion process here is not characterized by a decreasing sequence of thresholds on aspiration level.\footnote{For example, consider a setting with three individuals and three groups $N_1,N_2,N_3$, where $H_{N_2}\approx H_{N_3}$. Suppose $s_{N_1},s_{N_3}\approx 1$ and $s_{N_2}\approx 0$. Thus individual in $N_2$ put weight $s_{N_1}$ close to 1 on the expereince of the new product, and weight close to 2 on the experience of the incumbent product; while in $N_2$ put weight $s_{N_1}$ close to 1 on the experience of the new product, and weight close to 2 on the experience of the incumbent product.} Nonetheless, the model remains tractable and we have the following results.

\begin{proposition}\label{prop:KOL}
	Consider two social network $(s_{N_k})_{k=1}^G$ and $(s'_{N_k})_{k=1}^G$ and their demand at each period $(D_c^t,D_n^t)$ and $(D_c^{'t},D_n^{'t})$. We have $D_n^t\subseteq D_n^{'t}$ if
	\begin{enumerate}
		\item $(s'_{N_k})_{k=1}^G=z(s_{N_k})_{k=1}^G$ where $z\in[1,\frac{1}{\max_k s_{N_k}}]$, or
		\item $N$ is large enough, $v_{Hi}$ is invariant across $i$, and $\exists k>k'$ such that\footnote{The condition that $N$ is big enough ensures that the diffusion is characterized by a decreasing sequence of thresholds on aspiration level, so that ``early adopters" are well defined by their high aspiration levels.} 
		\begin{equation*}
			N_k(s'_{N_k}-s_{N_k})=N_{k'} (s_{N_{k'}}-s'_{N_{k'}})>0.
		\end{equation*}
		and $s_{N_{k''}}=s'_{N_{k''}}$ for all $k''\neq k, k'$.
	\end{enumerate}
\end{proposition}
 
The first bullet point of Proposition~\ref{prop:KOL} shows that when the network ties are stronger, it improves the diffusion of the new product. Note that under this condition, the ties with individuals who have lower aspiration levels, i.e. who are less willing to experiment, are also stronger. The result thus shows that the diffusion of the new product is improved as long as the ties with early adopters are strengthened to a similar degree as the ties with late adopters. It is in contrast with contagion models where individuals adopt the new product only if more than a fixed proportion of their neighbor has adopted where strengthening ties with both early and late adopters does not help diffusion. Also note that this result holds only because the new product is superior to the incumbent product, i.e., $v_{Hi}>v_L$ for all $i$. It thus suggests that a better-connected network will help differentiate good and bad products, by boosting diffusion of the former and stalling diffusion of the latter.

\begin{figure}
	\begin{center}
		\includegraphics[width=5cm]{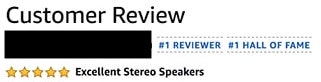}
	\end{center}
\caption{Amazon ranks reviewer and gives badges to top reviewers which makes their reviews more prominent. }
\label{amazon}
\end{figure}

The second bullet point shows that, unsurprisingly, when the society is more exposed to reviews/experiences of innovators, who are more inclined to experiment and switch to new products, other individuals would accumulate knowledge about the new product faster and thus switch to new products earlier. The more exposure of innovators could be driven by increased attention from the society based the perception that innovators are better equipped to evaluate new products, or platform strategy, e.g., see Figure~\ref{amazon}, that highlights reviews written by consumers that leave more (and more helpful) reviews. The result suggests that platform strategies that highlight reviews of individuals that frequently purchase new products would help diffusion of new innovation and thus increase the incentive of firms to innovate.

\section{Conclusion}
In this paper, I analyze a micro-founded model built on the case-based decision theory a la \cite{gilboa1995case,gilboa1996case,gilboa1997act}. The model is simple but captures a crucial element of product diffusion that individuals lack the knowledge to evaluate new products. Moreover, it includes rich model elements and allows analysis on the relationship between product diffusion and the characteristics of individuals, products and social network. For further studies, the model could be generalized and remain tractable to analyze other issues, including pricing and advertisement, related to innovation diffusion.

	\newpage
\bibliographystyle{econ}
\bibliography{reference_case}

\newpage
\appendix
\section{Proofs}

\subsection{Proof of Lemma~\ref{lemma_switch}}
\begin{proof}
	I prove by induction that individuals who choose the new product in period $t$ will choose the new product in period $t+1$ for all $t$, i.e., $D_{n}^{t}\subseteq D_{n}^{t+1}$ for all $t$. The Lemma then follows.
	
	First note that in period 1, only individuals in $N_{1}$ choose the new product: $U_{i}^1(p_{c})=\left((N-1)s+1\right)(v_{L}-H_{i})$ is strictly smaller than $U_{i}^1(p_n)=s_pU_{i}^1(p_{c})$ if and only if $H_i> v_L$, as $s_p\in(0,1)$. In period $t\geq 2$, for $i\in N_1$, 
	\begin{equation*}
		\begin{split}
				U_{i}^t(p_{n})=& s_p\sum_{t'=0}^{t-1}\sum_{j\in D_c^{t'}}s_{i,j} (v_L-H_{N_1})+\sum_{t'=0}^{t-1}\sum_{j\in D_n^{t'}} s_{i,j}(v_{Hj}-H_{N_1})\\
			&> \sum_{t'=0}^{t-1}\sum_{j\in D_c^{t'}}s_{i,j} (v_L-H_{N_1})\\&=U_{i}^t(p_{c})
		\end{split}
	\end{equation*}
	where the inequality is implied by $s_p\in(0,1)$ and $v_{Hj}\geq H_{N_1}>v_L$. Thus, individuals in $N_{1}$ choose new product in all period $t\geq 1$. $D_n^1\subseteq D_n^t$ for all $t\geq 1$.

	Now suppose $D_{n}^{t-1}\subseteq D_{n}^{t}$ for all $t=2,\cdots, T$, I proceed to prove $D_{n}^{T}\subseteq D_{n}^{T+1}$. First note that $U_{i}^t(p_{n})>U_{i}^t(p_{c})$ if and only if
	\begin{equation}\label{eq:buynewcondition}
		\begin{split}
			s_p\sum_{t'=0}^{t-1}\sum_{j\in D_c^{t'}}s_{i,j} (v_L-H_{i})+\sum_{t'=0}^{t-1}\sum_{j\in D_n^{t'}} s_{i,j}(v_{Hj}-H_{i})&> \sum_{t'=0}^{t-1}\sum_{j\in D_c^{t'}}s_{i,j} (v_L-H_{i});\\
\sum_{t'=0}^{t-1}\sum_{j\in D_n^{t'}} s_{i,j}(v_{Hj}-H_{i})&>(1-s_p)\sum_{t'=0}^{t-1}\sum_{j\in D_c^{t'}}s_{i,j} (v_L-H_{i});\\
\sum_{t'=0}^{t-1}\bigtriangleup_{t',n}(H_i)&>(1-s_p)\sum_{t'=0}^{t-1}\bigtriangleup_{t',c}(H_i)
		\end{split}
	\end{equation}
	where $\bigtriangleup_{t,n}(H_i)=\sum_{j\in D_n^{t}}s_{i,j}(v_{Hj}-H_i)$ and $\bigtriangleup_{t,c}(H_i)=\sum_{j\in D_c^{t}}s_{i,j}(v_L-H_i)$. Next, as $D_{n}^{T}=(D_{n}^{T}\setminus D_{n}^{T-1})\cup(D_{n}^{T-1}\setminus D_{n}^{T-2})\cup\cdots\cup(D_{n}^{2}\setminus D_{n}^{1})\cup D_{n}^{1}$, it suffices to prove $D_{n}^{t}\setminus D_{n}^{t-1}\subseteq D_{n}^{T+1}$ for $t=2,\cdots, T$.  I have already  proved above that $D_{n}^{1}\subseteq D_{n}^{T+1}$. For individuals in $D_{n}^{2}\setminus D_{n}^{1}$, as they choose the incumbent product in period $1$ but the new product in period $2$, it must be that the increase of the left-hand side of Inequality~\eqref{eq:buynewcondition} (from period $1$ to period $2$) is strictly greater than the increase of the right-hand side of the Inequality, i.e., $\bigtriangleup_{1,n}(H_i)>\bigtriangleup_{1,c}(H_i)$ for $i\in D_{n}^{2}\setminus D_{n}^{1}$.  Next, as $D_{n}^{1}\subseteq D_{n}^{t'}$ and $D_{c}^{t'}\subseteq D_{c}^{1}$ for all $t'=2,\cdots, T$ (by inductive assumption), for $i\in D_{n}^{2}\setminus D_{n}^{1}$ and all $t'=2,\cdots,T$,
	\begin{multline*}
		\bigtriangleup_{t',n}(H_i)=\sum_{j\in D_n^{t'}}s_{i,j}(v_{Hj}-H_i)\geq \sum_{j\in D_n^{1}}s_{i,j}(v_{Hj}-v_L)\\
		>\sum_{j\in D_c^{1}}s_{i,j}(1-s_p)(v_L-H_{i})\geq \sum_{j\in D_c^{t'}}s_{i,j}(1-s_p)(v_L-H_{i})=\bigtriangleup_{t',c}(H_i).
	\end{multline*}
    Combining with the fact that $\sum_{t'=0}^{1}\bigtriangleup_{t',n}(H_i)>\sum_{t'=0}^{1}\bigtriangleup_{t',c}(H_i)$ for $i\in D_{n}^{2}\setminus D_{n}^{1}$, we have $\sum_{t'=0}^{T}\bigtriangleup_{t',n}(H_i)>\sum_{t'=0}^{T}\bigtriangleup_{t',c}(H_i)$ and $D_{n}^{2}\setminus D_{n}^{1}\subseteq D_{n}^{T+1}$. Continuing the same procedure implies that individuals in $D_{n}^{t}\setminus D_{n}^{t-1}$ for $t=3,\cdots,T$ choose the new product in period $T+1$. The result follows.
\end{proof}

\subsection{Proof of Proposition~\ref{prop_segments}}
\begin{proof}
	I prove the first part of the Proposition by induction. In period $1$, only individuals with aspiration level $H_{i}> v_{L}$ consume the new product in period $1$, i.e., $\underline{H}_{1}=v_{L}$.
	
	Now suppose in period $t=2,\cdots,T$, there exists a unique threshold $\underline{H}_{t}$ such that only and all individuals whose aspiration level $H_{i}> \underline{H}_{t}$ consume the new product. Now given Lemma~\ref{lemma_switch}, in period~$T+1$, individuals with aspiration level strictly greater than $\underline{H}_{T}$ will continue to consume the new product. For an individual $i$ with $H_i\leq \underline{H}_{T}$, he chooses the new product in period~$T+1$ if and only if Inequality~\eqref{eq:buynewcondition} holds:
	\begin{equation*}
		\begin{split}
						\sum_{t'=0}^{T} \sum_{j\in D_n^{t'}}s(v_{Hj}-H_i)&>\sum_{t'=0}^{T} (1-s_p)\left[s(D_c^{t'}-1)+1\right](v_L-H_{i}).\\
		\end{split}
	\end{equation*}
	If $\sum_{t'=0}^{T} \sum_{j\in D_n^{t'}}s\geq \sum_{t'=0}^{T} (1-s_p)\left[s(D_c^{t'}-1)+1\right]$, all individuals consume $p_n$ as $v_{Hj}>v_L$ for all $j$, thus we set $\underline{H}_{T+1}=-\infty$ and the result follows. If in contrast $\sum_{t'=0}^{T} \sum_{j\in D_n^{t'}}s< \sum_{t'=0}^{T} (1-s_p)\left[s(D_c^{t'}-1)+1\right]$, the inequality then holds if and only if $H_{i}$ is big enough. Combing the two cases, there exists a unique threshold $\underline{H}_{T+1}$ such that only and all individuals whose aspiration level $H_{i}> \underline{H}_{T+1}$ consume the new product  in period $T+1$. The first part of the Proposition thus follows. The second part of the Proposition is immediately implied by Lemma~\ref{lemma_switch}.
\end{proof}

\subsection{Proof of Proposition~\ref{prop:coverage}}
\begin{proof}
	I first prove the ``if" statement of the first part of the Proposition. Suppose that there does not exist a $H_i\in[H_{N_{G}},H_{N_{1}})$ such that 
	\begin{equation*}
		\dfrac{\bar{v}_{>H_i}-H_i}{v_{L}-H_i}\leq \dfrac{(1-s_{p})\left[s[N-F(H_i)-1]+1\right]}{s[F(H_i)]}.
	\end{equation*}
	It implies that for all $H_i\in[H_{N_{G}},H_{N_{1}})$, 
	\begin{equation*}
		\begin{split}
			\dfrac{\bar{v}_{>H_i}-H_i}{v_{L}-H_i}&> \dfrac{(1-s_{p})\left[s[N-F(H_i)-1]+1\right]}{s[F(H_i)]}\\
			s\left[F(H_i)\right](\bar{v}_{>H_i}-H_i)&>(1-s_{p})\left[s(N-F(H_i)-1)+1\right](v_{L}-H_i)\\
			\sum_{j: H_j>H_i}s(v_{Hj}-H_i)&>\sum_{j: H_j\leq H_i} s_{i,j}\left(1-s_p\right)(v_L-H_i).
		\end{split}
	\end{equation*}
	That is, consider at period $t$ an individual with aspiration level $H_i$ who have yet consumed the new product, and suppose all other individuals with aspiration levels higher than $H_i$ consume the new product at period $t$. Then, given the consumption cases in period $t$, we have $\bigtriangleup_{t,n}(H_i)>\bigtriangleup_{t,c}(H_i)$. The same holds for all period $t'>t$. Thus, individual $i$ adopts the new product in some finite time $T=\ceil{\frac{U_{i}^{t}(p_c)-U_{i}^{t}(p_n)}{\bigtriangleup_{t,n}(H_i)-\bigtriangleup_{t,c}(H_i)}}$. Using induction arguments and the fact that individuals in $N_{1}$ adopt the new product in period~$1$, all individuals adopt $p_n$ in finite time.
	
	Now I prove the ``only if" statement.  Suppose to the contrary there exists some $H_i\in[H_{N_{G}},H_{N_{1}})$ such that
	\begin{equation*}
		\dfrac{\bar{v}_{>H_i}-H_i}{v_{L}-H_i}\leq \dfrac{(1-s_{p})\left[s[N-F(H_i)-1]+1\right]}{s[F(H_i)]}.
	\end{equation*}
	Also without loss suppose that $H_i\in[H_{N_{k}},H_{N_{k-1}})$ for some $k\in\lbrace2,\cdots, G\rbrace$, it implies that
	\begin{multline*}
					\dfrac{v_{>H_{N_{k}}}-H_{N_{k}}}{v_{L}-H_{N_{k}}}\leq \dfrac{v_{>H_{i}}-H_{i}}{v_{L}-H_{i}}\\
						\leq \dfrac{(1-s_{p})\left[s[N-F(H_i)-1]+1\right]}{s[F(H_i)]}=\dfrac{(1-s_{p})\left[s[N-F(H_{N_{k}})-1]+1\right]}{s[F(H_{N_{k}})]},
	\end{multline*}
	which in turn implies that, as $v_{L}>H_{N_{k}}$ and $v_{Hj}>v_L$ for all $j$,
	\begin{equation*}
		\begin{split}
			\sum_{j: H_j>H}s(v_{Hj}-H_{N_k})&\leq\sum_{j: H_j\leq H} s_{i,j}\left(1-s_p\right)(v_L-H_{N_k})
		\end{split}
	\end{equation*}
	for all $H\geq H_{N_k}$. Combined with the fact that only consumers in $N_{k'}$ where $k'<k$ would have adopted the new product before consumers in $N_{k}$, it implies that $\bigtriangleup_{t,n}(H_i)<\bigtriangleup_{t,c}(H_i)$ for $i\in N_k$ and for all $t$. Thus, consumers in $N_{k}$ never adopt the new product.

	For the second part of the Proposition, note that implied by the proof of the first part of the Proposition, individuals in $N_{\bar{G}}$ will not adopt the new product and individuals in $N_1,\cdots, N_{\bar{G}-1}$ will adopt the new product in finite time. Lastly, when $F(H_{i})$ increases in the sense of F.O.S.D., $v_{H>H_{i}}$ increases and  $\frac{(1-s_{p})\left[s[N-F(H_i)-1]+1\right]}{s[F(H_i)]}$ decreases, the last part of the Proposition follows.
\end{proof}

\subsection{Proof of Lemma~\ref{lemma:dominating}}
\begin{proof}
	In the following, I prove the Lemma in the case where $(v'_{Hi})_{i=1}^{N}>(v_{Hi})_{i=1}^{N}$ and $s'_p=s_p$ by induction. The cases where $(v'_{Hi})_{i=1}^{N}=(v_{Hi})_{i=1}^{N}$ and $s'_p>s_p$, and $(v'_{Hi})_{i=1}^{N}>(v_{Hi})_{i=1}^{N}$ and $s'_p>s_p$ follow similar arguments. At $t=1$, $\underline{H}_{1}=\underline{H}'_{1}=v_{L}$ and $D_n^{1}=D_n^{'1}=N_1$. At $t=2$,
	\begin{equation*}
		\begin{split}
			\sum_{j\in N_1}s(v_{Hj}-H_i)&\leq \sum_{j\in N_1}s(v'_{Hj}-H_i),\\
			(1-s_p)\left[s\left( 2N-N_1-2\right)+2\right](v_L-H_i)&=(1-s'_p)\left[s\left( 2N-N_1-2\right)+2\right](v_L-H_i).
		\end{split}
	\end{equation*}
	Thus, individuals who adopt the product $((v_{Hi})_{i=1}^{N},s_p)$ will also adopt the product $((v'_{Hi})_{i=1}^{N},s'_p)$, but the reserve is not true, i.e., $\underline{H}_{2}\geq\underline{H}'_{2}$. 
	
	Now suppose $\underline{H}_{t}\geq\underline{H}'_{t}$ for all $t=3,\cdots,T$. It implies that for $t=1,2,\cdots,T$
	\begin{equation*}
		D^{'t}_{c}\subseteq D^{t}_{c} \text{ and } D^{t}_{n}\subseteq D^{'t}_{n}.
	\end{equation*}
	We thus have
	\begin{equation*}
		\begin{split}
			\sum_{t'=0}^{T} \sum_{j\in D_n^{t'}}s(v_{Hj}-H_i)&\leq\sum_{t'=0}^{T} \sum_{j\in D_n^{'t'}}s(v_{Hj}-H_i)\\
			&\leq\sum_{t'=0}^{T} \sum_{j\in D_n^{'t'}}s(v'_{Hj}-H_i)
		\end{split}
	\end{equation*}
	and
		\begin{equation*}
		\begin{split}
			\sum_{t'=0}^{T} (1-s_p)\left[s\left( D_c^{t'}-1\right)+1\right](v_L-H_i)&\geq\sum_{t'=0}^{T} (1-s_p)\left[s\left( D_c^{'t'}-1\right)+1\right](v_L-H_i)\\
			&=\sum_{t'=0}^{T} (1-s'_p)\left[s\left( D_c^{'t'}-1\right)+1\right](v_L-H_i).
		\end{split}
	\end{equation*}
	The individuals who adopt the product $((v_{Hi})_{i=1}^{N},s_p)$ will also adopt the product $((v'_{Hi})_{i=1}^{N},s'_p)$, but the reserve is not true, i.e., $\underline{H}_{T+1}\geq\underline{H}'_{T+1}$. 
The result follows.
\end{proof}

\subsection{Proof of Proposition~\ref{prop:speedvsacceleration}}
\begin{proof}
	First note that $\underline{H}_{T+1}$ is the solution of
\begin{equation*}
	\begin{split}
		\dfrac{1}{1-s_{p}}\dfrac{v_{H}-\underline{H}_{T+1}}{v_{L}-\underline{H}_{T+1}}=\frac{s\sum_{t=0}^T (D_c^t-1)+T+1}{s\sum_{t=0}^T D_{n}^{t}}.
	\end{split}
\end{equation*}
Define $\lambda_{T}=\frac{s\sum_{t=0}^T (D_c^t-1)+T+1}{s\sum_{t=0}^T D_{n}^{t}}$, $\mathscr{F}(H)=\frac{1}{1-s_{p}}\frac{v_{H}-H}{v_{L}-H}$ and $\mathscr{F}'(H)=\frac{1}{1-s'_{p}}\frac{v'_{H}-H}{v_{L}-H}$. I first prove that $\mathscr{F}$ and $\mathscr{F}'$ cross at most once for all $H\leq v_{L}$, which will be useful for the rest of the proof. To see that, note that the first derivative of $\mathscr{F}-\mathscr{F}'$ w.r.t.~$H$ equals
\begin{equation*}
	\frac{\partial (\mathscr{F}-\mathscr{F}')}{\partial H}=\frac{1}{v_{L}-H}\left(\frac{v_{H}-v_{L}}{1-s_{p}}-\frac{v'_{H}-v_{L}}{1-s'_{p}}\right)
\end{equation*}
which is monotonic for all $H<v_{L}$ which implies that $\mathscr{F}$ and $\mathscr{F}'$ cross at most once for all $H\leq v_{L}$. 

I now proceed to prove the first part of the proposition. First note that $\mathscr{F}$ and $\mathscr{F}'$ go to infinity when $H\rightarrow^{-} v_{L}$ and go to $\frac{1}{1-s_{p}}$ and $\frac{1}{1-s'_{p}}$ respectively as $H\rightarrow -\infty$. Thus, when $s_{p}> s'_{p}$, we have either $\mathscr{F}(H)>\mathscr{F}'(H)$ for all $H$ if $\frac{v_{H}-v_{L}}{1-s_{p}}\geq \frac{v'_{H}-v_{L}}{1-s'_{p}}$, or otherwise $\mathscr{F}(H)>\mathscr{F}'(H)$ if and only if $H<\tilde{H}$ for some $\tilde{H}\in(-\infty,v_{L})$. I first prove the result for the latter case by induction. First, $\underline{H}_{2}< \underline{H}'_{2}$ implies $\mathscr{F}(H)>\mathscr{F}'(H)$ for all $H\leq \underline{H}'_2$. Next, $\lambda_2\geq \lambda'_2$ such that $\underline{H}_{3}=-\infty$ if $\underline{H}'_{3}=-\infty$. Now suppose $\underline{H}_{3},\underline{H}'_{3}>-\infty$ and denote $\underline{H}_{3}(\lambda_2)$ and $\underline{H}'_{3}(\lambda_2)$ as the solution of $\mathscr{F}(H)=\lambda_2$ and $\mathscr{F}'(H)=\lambda_2$ respectively, we have 
\begin{equation*}
	\underline{H}_{3}(\lambda_{2})\leq \underline{H}_{3}(\lambda'_{2})< \underline{H}'_{3}(\lambda'_{2})
\end{equation*}
where the first inequality is implied by $\lambda_{2}\leq \lambda'_{2}$ and that $\mathscr{F}(H)$ increases in $H$, while the second inequality is implied by $\mathscr{F}(H)>\mathscr{F}'(H)$ for all $H\leq \underline{H}'_2$ and that $\underline{H}_3,\underline{H}'_3\leq\underline{H}'_2$. 

Now suppose $\underline{H}_{t}\leq \underline{H}'_{t}$ for $t=3,\cdots, T$. Note that it implies that $\lambda_{t}\leq \lambda'_{t}$ for $t=3,\cdots, T$, and $\underline{H}_{T+1}=-\infty$ if $\underline{H}'_{T+1}=-\infty$. Now suppose $\underline{H}_{T+1},\underline{H}'_{T+1}>-\infty$, and denote $\underline{H}_{T+1}(\lambda)$ and $\underline{H}'_{T+1}(\lambda)$ as the solution of $\mathscr{F}(H)=\lambda$ and $\mathscr{F}'(H)=\lambda$ respectively, similar to the proof that  $\underline{H}_{3}<\underline{H}'_{3}$, we have 
\begin{equation*}
	\underline{H}_{T+1}(\lambda_{T})\leq \underline{H}_{T+1}(\lambda'_{T})< \underline{H}'_{T+1}(\lambda'_{T}).
\end{equation*}
and the result follows. Similar arguments also prove that $\underline{H}_{t}< \underline{H}'_{t}$ for all $t\geq 3$ when $\mathscr{F}(H)>\mathscr{F}'(H)$ for all $H$.

I now prove the second part of the proposition. Note that using similar arguments in the proof of the first part, if there exists some $t$ such that $\underline{H}'_{t}<\underline{H_{t}}$, we should have $\underline{H}'_{t'}<\underline{H_{t'}}$ for all $t'> t$. It implies that when $s_{p}<s'_{p}$ and $\underline{H}_{2}<\underline{H}'_{2}$, we either have case 2(a) or case 2(b) in the proposition. In the following, I provide in the following examples for the two cases 2(a) and 2(b). First, when $\underline{H}_{2}<\underline{H}'_{2}$ and $\underline{H}_{3}=-\infty<\underline{H}'_{3}$ (for example when $D_n^2$ is large), case 2(a) obviously hold. Second, when $\underline{H}_{2}+\epsilon=\underline{H}'_{2}>-\infty$ for some very small $\epsilon$, $D_n^t=D_n^{'t}$ and $\underline{H}_{3}>-\infty$, case 2(b) holds.
\end{proof}


\subsection{Proof of Proposition~\ref{prop:homophily}}
\begin{proof}
	Similar arguments in Lemma~\ref{lemma_switch} and Proposition~\ref{prop_segments} show that the diffusion pattern is characterized by a decreasing sequence of threshold $\underline{H}_{t}$, which is the solution (if it exists, and otherwise equals to $-\infty$) of
	\begin{equation}\label{eq:homophily}
		\frac{1}{1-s_{p}}\frac{s\sum_{t'=0}^{t-1}\sum_{j\in D_{n}^{t'}}(v_{Hj}-\underline{H}_{t})}{v_{L}-\underline{H}_{t}}=s\left[\sum_{t'=0}^{t-1}\left(\gamma(\frac{N}{G}-1)+ (D_{c}^{t'}-\frac{N}{G})\right)\right]+t-1.
	\end{equation}
	
	I prove the Proposition by induction. First, note that $D_{c}^{0}$, $D_{c}^{1}$ and $D_{n}^{1}$ do not change with $\gamma$. As the left hand side of Equation~\eqref{eq:homophily} increases in $\underline{H}_{t}$ and the right hand side increases in $\gamma$, $\underline{H}_{2}$ increases in $\gamma$. Now suppose $\underline{H}_{t}$ increases in $\gamma$ for all $t=3,\cdots, T$. First, it implies that $D_{c}^{t}$ increases in $\gamma$ and $D_{n}^{t}$ decreases in $\gamma$ for all $t=3,\cdots, T$, and the right hand side of Equation~\eqref{eq:homophily} increases in $\gamma$ and the left-hand side decreases in $\gamma$.
	Moreover, as the left hand side of Equation~\eqref{eq:homophily} increases $\underline{H}_{T+1}$, $\underline{H}_{T+1}$ increases in $\gamma$. The result follows.
\end{proof}

\subsection{Proof of Proposition~\ref{prop:KOL}}
\begin{proof}
	I first prove the first bullet point of the Proposition. First recall that we have for social network $(s_{N_k})_{k=1}^G$, $U_i^t(p_n)>U_i^t(p_c)$ if and only if
	\begin{equation*}
		\begin{split}
			\sum_{t'=0}^{T} \sum_{j\in D_n^{t'}}s_j(v_{Hj}-H_i)&>\sum_{t'=0}^{T} (1-s_p)\left[s_j(D_c^{t'}-1)+1\right](v_L-H_{i})\\
			\sum_{t'=0}^{T} \sum_{j\in D_n^{t'}}s'_j(v_{Hj}-H_i)&>\sum_{t'=0}^{T} (1-s_p)\left[s'_j(D_c^{t'}-1)+z\right](v_L-H_{i})\\
			&>\sum_{t'=0}^{T} (1-s_p)\left[s'_j(D_c^{t'}-1)+1\right](v_L-H_{i})
		\end{split}
	\end{equation*}
	Thus, fixing all other other parameters and if $(D_n^t)_{t=1}^T=(D_n^{'t})_{t=1}^T$, if and individual adopts the new product with social network $(s_{N_k})_{k=1}^G$, he also adopts the new product with social network $(s'_{N_k})_{k=1}^G$. The result follows with inductive arguments similar to the proof of Lemma~\ref{lemma:dominating}.
	
	Now I prove the second bullet point of the Proposition. First, note that when $N$ is large, the impact of self-connection can be ignored and thus $U_i^t(p_n)>U_i^t(p_c)$ if and only if
	\begin{equation*}
		\begin{split}
			\sum_{t'=0}^{t-1}\sum_{j\in D_{n}^{t'}}s_j (v_{H}-H_i)
			&=(1-s_p)\sum_{t'=0}^{t-1}\sum_{j\in D_{c}^{t'}}s_j(v_{L}-H_i).
		\end{split}
	\end{equation*}
	Then, similar arguments in Lemma~\ref{lemma_switch} and Proposition~\ref{prop_segments} show that the diffusion pattern is characterized by a decreasing sequence of threshold $\underline{H}_{t}$, which is the solution (if it exists, and otherwise equals to $-\infty$) of
	\begin{equation}\label{eq:KOL}
		\begin{split}
		\frac{1}{1-s_p}\frac{v_{H}-\underline{H}_{t}}{v_{L}-\underline{H}_{t}}=\frac{\sum_{t'=0}^{t-1}\sum_{j\in D_{c}^{t'}}s_j}{\sum_{t'=0}^{t-1}\sum_{j\in D_{n}^{t'}}s_j }
		\end{split}
	\end{equation}
	Now consider two social networks $(s_{N_k})_{k=1}^G$ and $(s'_{N_k})_{k=1}^G$ where $\exists k>k'$ such that
		\begin{equation*}
			N_k(s'_{N_k}-s_{N_k})=N_{k'} (s_{N_{k'}}-s'_{N_{k'}})>0.
		\end{equation*}
		and $s_{N_{k''}}=s'_{N_{k''}}$ for all $k''\neq k, k'$. First, in periods where individuals in both group $k$ and $k'$ adopt the incumbent product, the right hand-side of Equation~\eqref{eq:KOL} and thus the solution $\underline{H}_t$ are the same for the two social networks. In contrast, in periods where when $k$ adopts the new product and $k'$ adopts the new product, the right  hand-side of Equation~\eqref{eq:KOL} is smaller for social network $(s'_{N_k})_{k=1}^G$. As the left hand-side of Equation~\eqref{eq:KOL} increases in $\underline{H}'_t$, we have $\underline{H}'_t\leq \underline{H}_t$. The result thus follows.

\end{proof}

\end{document}